## Title: Ferromagnetism and Topology of the Higher Flat Band in a Fractional Chern Insulator

Authors: Heonjoon Park[1†], Jiaqi Cai[1†], Eric Anderson[1†], Xiao-Wei Zhang[2†], Xiaoyu Liu[2], William Holtzmann[1], Weijie Li[1], Chong Wang[2], Chaowei Hu[1], Yuzhou Zhao[1,2], Takashi Taniguchi[3], Kenji Watanabe[4], Jihui Yang[2], David Cobden[1], Jiun-haw Chu[1], Nicolas Regnault[5,6], B. Andrei Bernevig[5,7,8], Liang Fu[9], Ting Cao[2], Di Xiao[1,2*], Xiaodong Xu[1,2*]

[1]Department of Physics, University of Washington, Seattle, WA, USA
[2]Department of Materials Science and Engineering, University of Washington, Seattle, Washington, USA
[3]Research Center for Materials Nanoarchitectonics, National Institute for Materials Science, 1-1 Namiki, Tsukuba 305-0044, Japan
[4]Research Center for Electronic and Optical Materials, National Institute for Materials Science, 1-1 Namiki, Tsukuba 305-0044, Japan
[5]Department of Physics, Princeton University, Princeton, New Jersey 08544, USA
[6]Laboratoire de Physique de l'Ecole Normale Supérieure, ENS, Université PSL, CNRS, Sorbonne Université, Université Paris-Diderot, Sorbonne Paris Cité, 75005 Paris, France
[7]Donostia International Physics Center, P. Manuel de Lardizabal 4, 20018 Donostia-San Sebastian, Spain
[8]IKERBASQUE, Basque Foundation for Science, Bilbao, Spain
[9]Department of Physics, Massachusetts Institute of Technology, Cambridge, Massachusetts 02139, USA
[†] These authors contributed equally to the work.
Correspondence to: dixiao@uw.edu, xuxd@uw.edu

**Abstract: The recent observation of the fractional quantum anomalous Hall effect in moiré fractional Chern insulators provides an opportunity to investigate zero magnetic field anyons. To potentially realize non-Abelian anyons, one approach is to engineer higher flat Chern bands that mimic higher Landau levels. Here, we investigate the interaction, topology, and ferromagnetism of the second moiré miniband in twisted $MoTe_2$ bilayers. At half filling of the second miniband, we observe spontaneous ferromagnetism and an incipient Chern insulator state. The Chern numbers of the top two moiré flat bands exhibit opposite signs for twist angles above 3.1°, but share the same sign near 2.6°, consistent with theoretical predictions. In the 2.6° device, increasing magnetic field induces a topological phase transition via band crossing between opposite valleys, resulting in an emergent state with Chern number $C$ = -2. Additionally, an insulating state at half filling of the second valley-polarized band suggests a charge-ordered state is favored over the fractional Chern insulator state. These findings lay a foundation for understanding the higher flat Chern bands, crucial for the discovery of non-Abelian fractional Chern insulators.**

### Main text:

The fractional Chern insulators (FCIs)[1-6] are a family of topologically ordered states that are zero-field lattice analogues of the fractional quantum Hall (FQH) states. Their lattice-based nature suggests that they may be more robust than the FQH states, allowing them to exist without a magnetic field at, hopefully, elevated temperatures. Recent observations of the fractional quantum anomalous Hall (FQAH) effect in twisted $MoTe_2$ bilayer ($tMoTe_2$)[7-12] and pentalayer rhombohedral graphene aligned with hexagonal boron nitride[13] have indeed confirmed the existence of these FCI states at zero field and their persistence up to 2 K in the case of $tMoTe_2$[9]. All identified zero-field FCI states to date belong to the Jain sequence of their cousin FQH states[7-10,13], and are expected to host abelian anyons. This observation is consistent with the picture of the first flat Chern band of $tMoTe_2$ being somewhat analogous to the lowest Landau level (LL)[14-24], but with important differences including the absence of the 1/3 state, band mixing effects, and the landscape of spontaneous magnetization.

The first moiré band of $tMoTe_2$ carries valley-opposite Chern numbers (Fig. 1a-b) originating from the layer pseudospin skyrmion lattice[15,25]. Theoretical studies have revealed that moiré bands at higher energy also exhibit twist angle dependent band topology[26-29]. Intriguingly, a sequence of consecutive Chern bands has been predicted to exist in a certain range of the twist angles[26]. The higher Chern bands, like higher LLs, differ in effective interaction and quantum geometry from the first flat band, despite sharing the same Chern number. Exploring these higher moiré flat bands may yield even-denominator FCI states[30-37], by exploiting the analogy with higher LLs in the FQH framework. Band-projected calculations without band mixing have indeed pointed out the possibility of non-Abelian FCIs in the second moiré Chern band at half-filling[38-41], akin to the non-Abelian Moore-Read state in the half-filled second LL. However, the physics of $tMoTe_2$ is strongly affected by band mixing[42-44], which is difficult to treat theoretically. As such, experimental studies of the higher Chern bands are essential to guide modeling and understanding of their magnetic and topological character.

Here, we report the observation of robust spontaneous ferromagnetism of the second Chern band, with twist-angle dependent topology, in $tMoTe_2$. The devices used in the main figures have two twist angles ($\theta$) of approximately 2.6° and 3.8°. Additional data from devices with $\theta$ ranging between 2.6-3.8° are provided in Extended Data Fig.1. Two types of device were employed: simple structures, suitable for detection of ferromagnetism using reflective magnetic circular dichroism (RMCD, Ref.[7,45]); and Hall bars with additional fabrication steps to improve electrical contacts[9], suitable for magnetotransport. The low-temperature contact resistance was typically 15-30 kΩ (see Methods and Extended Data Fig. 2 for details).

**Spontaneous ferromagnetism near $\nu$ = -3**

We first consider RMCD measurements of the spontaneous ferromagnetism in a 2.6° $tMoTe_2$ device. Figure 1c is a plot of RMCD signal versus filling factor $\nu$ (defined as minus the number of holes per moiré unit cell) and perpendicular electric field $D/\varepsilon_0$ at a temperature of $T$ = 1.6 K and under a small magnetic field $\mu_0 H$ = 50 mT (see Methods). The role of the small magnetic field is to suppress magnetic fluctuations which produce an unstable RMCD signal while sweeping the hole density at zero field, as illustrated in Extended Data Fig. 1a. Consistent with prior reports[7,8,45], the RMCD signal is pronounced near $\nu$ = -1, where the first non-interacting Chern band in each valley is half-filled and exhibits spontaneous valley polarization with broken time reversal symmetry. The top two panels in Fig. 1d show the RMCD signal obtained by cycling the magnetic field at $\nu$ = -1 and -2/3. Hysteresis loops confirm the presence of ferromagnetic order. Near $\nu$ = -2, both the RMCD signal and the hysteresis vanish. This implies that the flat Chern bands for both $\pm K$ valleys are fully filled and energy degenerate, resulting in the absence of spontaneous spin/valley polarization.

Near $\nu$ = -3, where the second moiré band is half-filled, an RMCD signal appears, indicating spontaneous ferromagnetism. This is confirmed by hysteresis seen upon cycling $\mu_0 H$, as shown in the bottom panel in Fig. 1d (see also Extended Data Fig. 3). The signal persists up to moderate electric fields and exhibits "wings" in the $\nu$-$D/\varepsilon_0$ plane that resemble those seen near $\nu$ = -1, suggesting a common underlying mechanism. The ferromagnetism $\nu$ = -3 is robust for the entire angle range of 2.6° to 3.8° as seen from dual gate maps of RMCD signal in Extended Data Fig. 1. A prior report[45] has shown that near $\nu$ = -1, the ferromagnetism arises from direct exchange interactions between the doped holes. This is also the case for the second moiré miniband, as shown by the calculated hole density distribution within the moiré unit cell (Extended Data Fig. 4). As the strength of the perpendicular electric field increases, the system becomes layer-polarized, and the interplay between on-site Columb repulsion and interlayer charge transfer energy determines the ferromagnetic phase diagram, leading to the wing-like features.[45]

We plot the difference in RMCD signal between $\mu_0 H$ swept down and up (ΔRMCD) as a function of $\nu$ in Fig. 1e. The coercive field is enhanced at fillings $\nu$ = -1 and -2/3, which correspond to integer and fractional Chern insulator states, consistent with previous findings[7,45]. In contrast, the coercivity remains around 25 mT with minimal variation near $\nu$ = -3, suggesting a weaker magnetic stability compared to $\nu$ = -1 and -2/3. Temperature dependent RMCD measurements yield Curie temperatures $T_c$ of approximately 10 K at $\nu$ = -1 (Fig. 1f) and 6 K at $\nu$ = -3 (Fig. 1g; see also Extended Data Fig. 3). The smaller $T_c$ at $\nu$ = -3 also indicates a weaker magnetic interaction strength compared to $\nu$ = -1, possibly due to stronger screening at higher doping, larger bandwidth (smaller density of states), or stronger band mixing due to closer nearby bands. In addition, we note that the $T_c$ at $\nu$ = -1 is smaller than the value of ≈14 K seen in larger twist-angle ($\theta \approx$ 3.6°) devices[45]. This can be attributed to the weaker exchange interactions at smaller $\theta$ due to the larger moiré period.

**Anomalous Hall effect near $\nu$ = -3**

We next investigate the topological properties of the second moiré band using electrical transport measurements. Figures 2a and b show the longitudinal resistance ($R_{xx}$) and Hall resistance ($R_{xy}$) as a function of $\nu$ and $D/\varepsilon_0$ for a $\theta \approx$ 2.6° device at $T$ = 15 mK. Figures 2c and d show the temperature dependence of $R_{xx}$ and $R_{xy}$ vs $\nu$ at a fixed electric field $D/\varepsilon_0$ = 0. To mitigate slight mixing between $R_{xx}$ and $R_{xy}$ and magnetic fluctuations, $R_{xx}$ ($R_{xy}$) is symmetrized (antisymmetrized) with respect to a finite magnetic field of ±0.1 T for Figs. 2a and b and ±0.5 T for Figs. 2c and d. Near $\nu$ = -1, the almost vanishing $R_{xx}$ accompanied by a plateau of $R_{xy}$ at $\pm h/e^2$ (see Fig. 2e) signifies a well-developed integer QAH state. At $\nu$ = -2/3, a dip in $R_{xx}$ together with $R_{xy} \approx -3h/2e^2$ is consistent with a -2/3 FCI state (Figs. 2c, d; see also Extended Data Fig. 5), which is somewhat obscured by the higher contact resistance at very low doping. These signatures of the QAH state at filling factor -1 and FCI state at -2/3 are in line with previous reports[9,10].

An anomalous Hall signal with hysteresis in $R_{xy}$ is also observed near $\nu$ = -3 (Fig. 2f), where a minimum is seen in $R_{xx}$ at ≈ 7 kΩ (Fig. 2c) accompanied by a maximum in $R_{xy}$ at ≈ -12 kΩ (Fig. 2d). The large but non-quantized value of $R_{xy}$ and dip in $R_{xx}$ hint at an incipient Chern insulator state with the Chern gap not fully developed. Indeed, in Figs. 2c and d it can be seen that, as $T$ increases from 50 mK to 6 K, $R_{xx}$ at $\nu$ = -3 initially increases while $R_{xy}$ drops (but remains finite). Then, above the Curie temperature, $T_c \approx$ 6 K, $R_{xx}$ begins to decrease. A qualitatively similar but much stronger temperature dependence is seen in the QAH state at $\nu$ = -1 (Extended Data Fig. 5).

These observations support the band structure proposed in Fig. 1b (see also Extended Data Fig. 4). At $\nu$ = -1, the exchange coupling between the two spin-valley bands is strong enough to produce a full gap between the two narrow Chern bands, forming a QAH state. However, due to the larger bandwidth at $\nu$ = -3 (Extended Data Fig. 4a), there is still some overlap of the Chern bands. This picture is consistent with our Hartree-Fock calculations (Extended Data Fig. 4b, see Methods for details). At $\nu$ = -2, $R_{xy}$ vanishes at zero magnetic field (see Extended Data Fig. 5e) indicating equal occupation of the +K and -K valleys. With increasing magnetic field, $R_{xy}$ rapidly rises and saturates around 7-8 kΩ. Further investigation is needed to fully understand the nature of this state.

**Twist-angle dependent anomalous Hall effect**

Figure 3a illustrates the hysteretic component of $R_{xy}$, denoted by $\Delta R_{xy}$, as a function of the magnetic field and filling from a device with $\theta$ = 3.8°. Consistent with previous reports[7], enhancements in the coercive field are seen at $\nu$ = -1, $\nu$ = -2/3, and $\nu$ = -3. The most remarkable aspect of the data, however, is the opposite sign of the anomalous Hall effect at $\nu$ = -3 and at $\nu$ = -1, which can also be seen in the selected magnetic field sweeps shown in Figs. 3b and c. After the magnetization is initialized with a positive magnetic field, at $\nu$ = -1, $R_{xy}$ is negative (and close to -$h/e^2$) as in the 2.6° device above. At $\nu$ = -3, for the same sign of magnetic field, $R_{xy}$ has the opposite sign (≈ +7 kΩ) in this $\theta$ = 3.8° device, while it is negative (≈ -12 kΩ) in the 2.6° device. A direct comparison of the filling factor dependence of the anomalous Hall effect in both devices is shown in Extended Data Fig. 6.

One plausible explanation for the sign reversal of the anomalous Hall signal at $\nu$ = -3 is a change in the topology of the second band as the twist angle varies. Indeed, several theoretical works have predicted a twist-angle dependent valley Chern number for the second band[17,26,28]. Large-scale density functional theory (DFT) calculations of the band structure at the commensurate twist angles of 2.45° and 3.89°, chosen to be close to the experimental values, are shown in Fig. 3d. For a given valley, the Chern number of the first Chern band at the $K$ (spin up) valley remains at $C$ = +1 for both angles, while the second band has $C$ = +1 at 2.45° and $C$ = -1 at 3.89°. Our DFT calculations[26] predict that the Chern number transition occurs for $\theta_c$ near 2.6°. An additional sample with $\theta \approx 3.1°$ also shows opposite signs for the first and second Chern bands, i.e. (+1, -1) (Extended Data Fig. 6), experimentally constraining $\theta_c$ to lie between 2.6° and 3.1°.

This predicted transition is linked to the evolution of the surface moiré potential as a function of twist angle, caused by the competition between piezoelectric and ferroelectric effects[26]. The electrostatic Hartree potential drop $\Delta v_H$ between the top and bottom layer[26] is plotted in Fig. 3e. At the smaller $\theta$ of 2.45°, the maxima (minima) of $\Delta v_H$ are located between the MX (XM) and MM sites, resulting in the electron wavefunction distribution within the moiré unit cell as seen in Extended Data Fig. 4. In contrast, at the larger $\theta$ of 3.89°, the maximum lies on the MX site, while the minimum lies on the XM site. This shift in the Hartree potential extrema as a function of twist angle causes the change in band topology.

**Chern insulator from Zeeman-field induced band crossing**

A direct method to experimentally determine the band topology of the second moiré band is to measure the evolution of features with magnetic field, which reveals the Chern number of gapped states via the Streda formula[7,46-48]. Figures 4a, b and 4e, f show plots ("fan diagrams") of symmetrized $R_{xx}$ and antisymmetrized $R_{xy}$ vs $\mu_0 H$ and $\nu$ for the $\theta$ = 3.8° and 2.6° devices, respectively. The measurements were taken at $T < 30$ mK and $D/\varepsilon_0 = 0$. Similar measurements were also performed at large $D/\varepsilon_0$, where the appearance of multiple Landau levels allows us to accurately extract the carrier density and assign filling factors (Extended Data Fig. 7).

The fan diagram reveals a number of electronic phases, illustrated in the Wannier diagrams in Figs. 4c and g. The states of interest are annotated with the Streda slope and intercept ($C$, $\nu$), with $C$ the Chern number and $\nu$ the filling factor. The dispersive features for the QAH (-1, -1) and FCI (-2/3, -2/3) states are consistent with prior reports[7,9,10]. The $\nu = -3$ state, in the 3.8° device exhibits a slope of $C = +1$, with $R_{xy} \approx h/e^2$ and nearly vanishing $R_{xx}$ at high magnetic field (Fig. 4d). In contrast, in the 2.6° device, the $\nu = -3$ slope is $C = -1$. This is consistent with the inferred topology of the second Chern band and its dependence on twist angle previously discussed, as the Streda slope measures the sum of the Chern numbers up to the Fermi level.

For the 2.6° device, magnetic field-induced phases emerge across a wide range of filling factors centered around $\nu = -2$ (Fig. 4e-g). At low magnetic fields (< 7 T) around $\nu = -2$, $R_{xx}$ is finite ($\approx$ 28 kΩ) with little dependence on $\mu_0 H$ while $R_{xy} \approx$ 7-8 kΩ. Around 7 T, an intriguing feature can be seen near the hole-doped side of $\nu = -2$, where $R_{xx}$ vanishes and $R_{xy}$ is close to $-h/2e^2$. Figure 4h displays traces of $R_{xx}$ and $R_{xy}$ versus $\nu$ at 13T. Plateaus with nearly vanishing $R_{xx}$ and $R_{xy}$ quantized to $-h/2e^2$ are clearly visible slightly above $\nu = -2$. A line with Streda slope corresponding to $C = -2$ and with an intercept of $\nu = -2$ at zero field aligns well with this dispersive feature. The data demonstrate a magnetic field induced $C = -2$ Chern insulator state.

We can understand this topological phase transition to result from a Zeeman-induced band crossing between opposite valleys. As shown schematically in Fig. 4i, the top valence bands of the $\pm K$ valleys are equally occupied at $\nu = -2$ at small magnetic field. The valleys have opposite Chern numbers, resulting in a net $C = 0$. With increasing magnetic field, the second flat band of the $-K$ valley shifts up in energy while the first flat band of the $+K$ moves down in energy. Above a critical magnetic field a band crossing occurs, resulting in two hole-filled flat bands from the same valley. Our measurements demonstrate that the Chern numbers of these first and second flat bands with the same valley index have the same sign, $C = +1$, for the $K$ valley in 2.6° $tMoTe_2$. This leads to a Chern insulator state with $C = -2$ at high magnetic field as observed. More detailed measurements suggest that the magnetic field-induced topological phase transition is a first-order phase transition (Extended Data Fig. 8) and the $C = -2$ state has an activation gap of around 13 K at 13 T (Extended Data Fig. 9).

The fan diagram also reveals a magnetic field induced insulating state at $\nu = -3/2$ above 7 T (Fig. 4j). This state can be clearly seen in Extended Data Fig. 7, where $R_{xx}$ is plotted versus $\nu$ and $D/\varepsilon_0$ at selected magnetic fields above $\mu_0 H = 7$ T. The state does not disperse with magnetic field. This observation implies that at half-filling of the second valley polarized Chern band after band crossing, a charge ordered insulating state is preferred over the FCI state at high magnetic fields. While possible non-Abelian FCI states have recently been proposed at $\nu = -3/2$[38-40,49-52], our observations show that under these experimental settings at a twist angle of 2.6°, a charge-ordered insulating state is likely the ground state.

The identification of ferromagnetism and an incipient Chern insulator at $\nu = -3$, along with the observation of a Zeeman field-induced Chern insulator phase transition, provide an excellent starting point for engineering even-denominator FCI states in higher Chern bands. Considering recent indications of a possible time-reversal symmetric fractional quantum spin Hall state at $\nu$ = -3 for a 2.1° twisted $MoTe_2$(Ref.[53]), our experiments provide strong constraints and guidance for both theoretical and experimental exploration of the correlated phases and topology in this rich system.

**Acknowledgements:** The authors thank Jiabin Yu and Ady Stern for insightful discussion. This work is mainly supported by DoE BES under award DE-SC0018171. Electrical transport measurements of QAH insulator are partially supported by AFOSR FA9550-21-1-0177. The understanding of magnetism and topological phase diagram is partially supported by AFOSR Multidisciplinary University Research Initiative (MURI) program, grant no. FA9550- 19-1-0390. The growth of the $MoTe_2$ crystal is supported by the Center on Programmable Quantum Materials, an Energy Frontier Research Center funded by DOE BES under award DE-SC0019443. The authors also acknowledge the use of the facilities and instrumentation supported by NSF MRSEC DMR-2308979. BAB was supported by Simons Investigator Grant No. 404513 and the Gordon and Betty Moore Foundation's EPiQS Initiative (Grant No. GBMF11070). EA acknowledges the support by the National Science Foundation Graduate Research Fellowship Program under Grant No. DGE-2140004. K.W. and T.T. acknowledge support from the JSPS KAKENHI (Grant Numbers 21H05233 and 23H02052) and World Premier International Research Center Initiative (WPI), MEXT, Japan. JHC and XX acknowledge support from the State of Washington funded Clean Energy Institute.

**Author contributions:** XX conceived and supervised the project. HP and EA fabricated the samples. HP and JC performed the transport measurements. EA performed the optical measurements, with support from WH and WL. JC, DHC, and XX provided dilution fridge measurement support. CH, YZ, JY, and JHC synthesized and characterized bulk $MoTe_2$ crystals. XWZ, TC, DX performed large-scale DFT calculations. XL, CW, TC, and DX performed Hartree-Fock calculations. HP, JC, EA, DHC, NR, BAB, LF, TC, DX, and XX analyzed and interpreted the results. TT and KW synthesized the hBN crystals. XX, HP, JC, DX, DHC, BAB, and LF wrote the paper with input from all authors. All authors discussed the results.

**Competing Interests:** The authors declare no competing financial interests.

**Methods:**

**Sample fabrication.** Single crystals of 2H-$MoTe_2$ were grown by the flux method, using Te as the self-flux. The high-quality crystals allow us to observe quantum oscillations starting from 2T upon gate tuning the device out of flat band regime, aiding us to accurately determine the moiré filling factor.

Low contact resistance in twisted 2D semiconductor devices was achieved by a standard contact gate and Pt contact scheme[9] (Extended Data Fig. 2a). In brief, mechanically exfoliated hBN**,** 2H-$MoTe_2$, and graphite flakes are found using an optical microscope and confirmed to be free of residue by atomic force microscopy (AFM) or contrast-enhanced optical microscopy. First, a standard polycarbonate-based dry transfer technique was used to pick up an hBN flake to serve as a bottom gate dielectric, followed by a graphite bottom gate electrode. The hBN-graphite stack was then released onto a silicon substrate. Hall bar contacts were defined by standard e-beam lithography, metal deposition (Pt, 8nm) was performed by e-beam evaporation, and the back gate was subsequently cleaned by an AFM in contact mode. We note that the homogeneity of the Pt film is crucial for achieving low contact resistance, and we typically aim for a root mean square height variation less than 300 pm. Second, in an Argon glovebox environment with $O_2$ and $H_2O$ levels < 0.1 ppm, $MoTe_2$ monolayers were found, confirmed to be clean, and cut using an AFM tip. Dry transfer was used to pick up a thin BN as contact gate dielectric, part of the $MoTe_2$ monolayer, and the other part of the $MoTe_2$ monolayer after rotating the stage to the desired angle. The stack was released onto the bottom gate and contacts after alignment. Third, the stack was washed off and cleaned by an AFM in contact mode in atmosphere, followed by metal deposition of a Pt (8nm) contact gate pattern and Cr/Au (5/70 nm) connecting electrodes, followed by another round of AFM contact mode cleaning. Finally, the top gate structure (hBN/graphite/hBN) was picked up using dry transfer and released onto the device. Optical and AFM images of the device can be seen in Extended Data Fig. 2. Optical devices for RMCD measurements were built by a similar method but without a contact gate, and sometimes a graphite electrode was used instead of a Pt electrode for grounding the device.

A persistent difficulty in transport studies of TMD moiré heterostructures has been the formation of Schottky barriers between the metallized TMD and the intrinsic TMD channel. The resulting high contact resistance introduces significant noise in transport measurements and poses a challenge to access the full electronic phase space, especially around zero electric field, where the majority of topological and strongly correlated phenomena occur in twisted TMD homobilayers. The above contact gate design allows us to locally hole-dope the boundary between the metallized TMD region and intrinsic TMD channel by applying a large negative voltage on the contact gates (Extended Data Figs. 2b-c), enabling the formation of near-ohmic contacts to $MoTe_2$ homobilayers down to $T = 10$ mK. In Extended Data Figs. 2f-g, we plot the two-terminal resistance between two contacts while floating all other contacts. Although this method does not provide an exact measure of contact resistance, it offers an upper bound of (approximately twice) the contact resistance. We observe that the contact resistance increases at low doping levels, likely due to intrinsic crystal defects, consistent with previous reports[9].

**Electrical measurements.** Transport measurements were conducted in a Bluefors dilution refrigerator with a 13.5 T magnet and a base phonon temperature of approximately 20 mK and electron temperature of 150 mK. Standard AC lock-in measurements were used to measure four-terminal $R_{xx}$ and $R_{xy}$ signals using a low ac current excitation of 0.2-0.5 nA.

**Reflective magnetic circular dichroism (RMCD) measurements.** RMCD measurements were performed in a closed-loop magneto-optical exchange gas cryostat (attoDRY 2100) with an attocube *xyz* piezo stage, a 9T out-of-plane superconducting magnet, and a base temperature of 1.6 K. RMCD excitation was achieved by filtering a broadband supercontinuum source (NKT SuperK FIANIUM FIU-15) via dual-passing through a monochromator. RMCD signal is proportional to the difference in reflectance of right and left circularly polarized (RCP and LCP) light normalized by the total reflectance. The excitation was chopped at 1 kHz and the polarization was alternated between RCP and LCP with a photoelastic modulator at 50 kHz. The reflected light was detected by an InGaAs avalanche photodiode and read out by two lock-in amplifiers (SR830) at 50 kHz and 1 kHz. This scheme allows for continuous detection of both the reflectance difference between RCP and LCP and the total reflectance, thus allowing RMCD to be extracted.

**Estimation of filling factor based on doping density.** The carrier density $n$ and electric field $D$ on the sample were derived from top (bottom) gate voltages $V_{tg}$ and $V_{bg}$ using $n = (V_{tg}C_{tg} + V_{bg}C_{bg})/e - n_{offset}$ and $D/\varepsilon_0 = (V_{tg}C_{tg} - V_{bg}C_{bg})/2\varepsilon_0 - D_{offset}$, where $e$ is the electron charge, $\varepsilon_0$ is the vacuum permittivity, and $C_{tg}$ and $C_{bg}$ are the top and bottom gate capacitances obtained from the gate thickness measured by AFM. $D_{offset} \approx 0$ was inferred from the dual gate resistance map. The high field Landau fan diagram was used to calibrate the capacitance of gates. A good correspondence between the capacitance model and the Landau fan diagram measurement was found. Moiré filling number was determined by tracking the Landau fan down to zero magnetic field. In optical measurements, moiré filling number was determined by photoluminescence measurement which shows the suppression of integrated spectral signal at integer and fractional fillings, as shown previously[7].

**Density functional theory calculation.** The lattice relaxations of moiré superlattices are addressed through the utilization of neural network (NN) potentials. These potentials are parameterized based on the deep potential molecular dynamics (DPMD) method[54,55]. Training datasets are obtained from 5000-step ab initio molecular dynamics (AIMD) simulations for a 6° $tMoTe_2$ at 500 K using the VASP package[56]. The van der Waals corrections are considered within the D2 formalism[57]. Further details regarding the parameterization of NN potentials can be found in Ref.[26]. The NN potentials are then employed to relax the moiré superlattices using the LAMMPS package[58] until the maximum atomic force is below $10^{-4}$ eV/Å.

The moiré mini band structures are calculated using the SIESTA package[59]. The optimized norm-conserving Vanderbilt (ONCV) pseudopotential[60], the Perdew-Burke-Ernzerhof (PBE) functional [61], and a double-zeta plus polarization basis are chosen. Spin-orbit coupling (SOC) is incorporated using the on-site approach[26]. To calculate the valley-resolved electron density and wave function, a small Zeeman field is added. A k-grid of 3×3×1 is used to sample the moiré Brillouin zone for density calculations. The interlayer Hartree potential drop is calculated as the difference between the potential above and below the moiré superlattice, using the same treatment as described in Ref.[26].

**Hartree-Fock calculation.** Starting with the DFT calculations of $tMoTe_2$ bilayer at 2.45°, we constructed three Wannier orbitals for each valley (spin) localized at the MX, XM and MM site, respectively (Extended Data Figs. 4h-j). Based on the constructed Wannier orbitals[62], we performed Hartree-Fock mean field calculations[38] at filling factor $\nu = -3$ on a 12×12 k-mesh. Since the Hartree-Fock method tends to overestimate the interaction effects, we reduce the interaction by selecting $\varepsilon = 60$. The obtained band structure is shown in Extended Data Fig. 4b. Future studies are required to incorporate detailed screening and band-mixing effects due to multiple nearby bands.

To understand why the second Chern band is not fully gapped by exchange coupling, we calculate the hole wavefunction for different points in momentum space (Extended Data Figs. 4d-g). For the second moiré miniband, the MBZ corner $\kappa$ points arise from the MX and XM moiré orbitals, which form a honeycomb lattice in real space. However, the MBZ high symmetry $\gamma$ point forms a triangular lattice with orbitals located at MM sites. This discrepancy leads to a Fock energy inhomogeneity across momentum space for the second miniband. The Fock energy splitting near the $\kappa$ point is larger than that of the $\gamma$ point as evidenced by the Hartree-Fock calculation in Extended Data Fig. 4b.

**Data Availability:** The datasets generated during and/or analyzed during this study are available from the corresponding author upon reasonable request.

**Methods-only References:**

**Figures:**

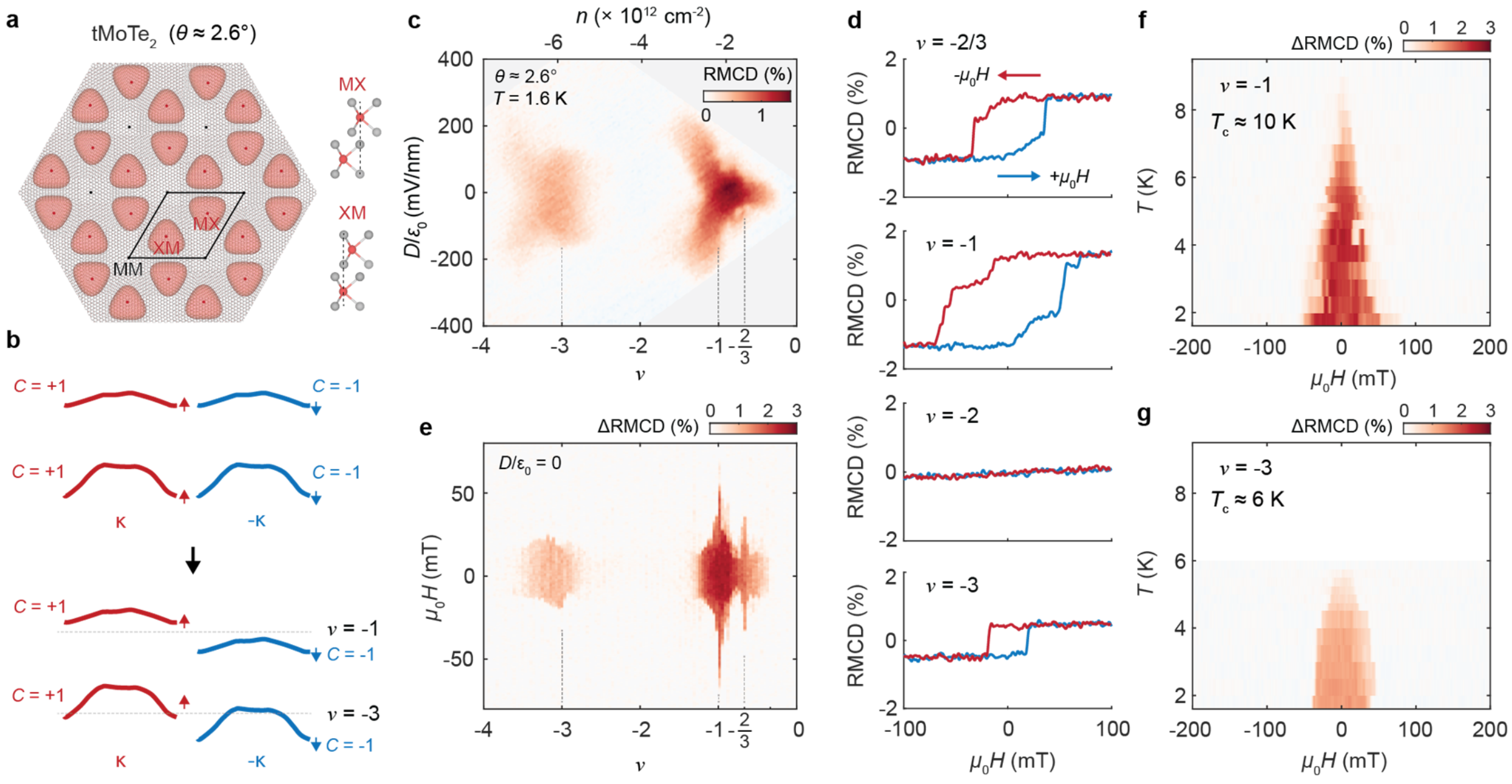

**Figure 1 | Spontaneous ferromagnetism in the second flat Chern band. a**, Schematic of the moiré superlattice of $tMoTe_2$ with a twist angle of 2.6°, with $C_3$ symmetric high symmetry sites MM, MX, and XM denoted. At half-filling of the flat Chern bands, the holes mainly occupy the MX and XM sites to form a honeycomb lattice. Direct exchange interaction between sites induces ferromagnetism. **b**, Top, Schematic of Chern bands in 2.6° $tMoTe_2$ without interactions. The first and second Chern bands have a spin/valley degeneracy, with opposite Chern numbers for +*K*/spin ↑ (red) and -*K*/spin ↓ (blue). Bottom, at half filling of the first or second Chern bands ($\nu = -1$ and $\nu = -3$), interaction-induced magnetic order breaks the spin/valley degeneracy. **c**, Reflective magnetic circular dichroism (RMCD) signal as a function of filling factor ($\nu$) and electric field ($D/\varepsilon_0$). A small magnetic field ($\mu_0 H$) of 50 mT is applied to suppress magnetic fluctuations. **d**, RMCD vs $\mu_0 H$ at selected $\nu$ as the field is swept down (red) and up (blue). Clear signatures of ferromagnetism are visible at $\nu = -2/3$, $\nu = -1$, and $\nu = -3$. Twist angle dependent RMCD data (see Extended Data Fig. 1) implies that the ferromagnetism is robust at $\nu = -3$ for the entire angle range from 2.6° to 3.8°. **e**, Hysteretic component of RMCD (ΔRMCD) vs $\nu$ and $\mu_0 H$. Large enhancements of the coercive field can be seen right at $\nu = -1$ and $\nu = -2/3$, while the coercive field remains roughly constant around $\nu = -3$. **f-g**, Temperature dependence of ΔRMCD for $\nu = -1$ (**f**) and $\nu = -3$ (**g**), indicating a Curie temperature ($T_c$) of about 10 K and 6 K, respectively.

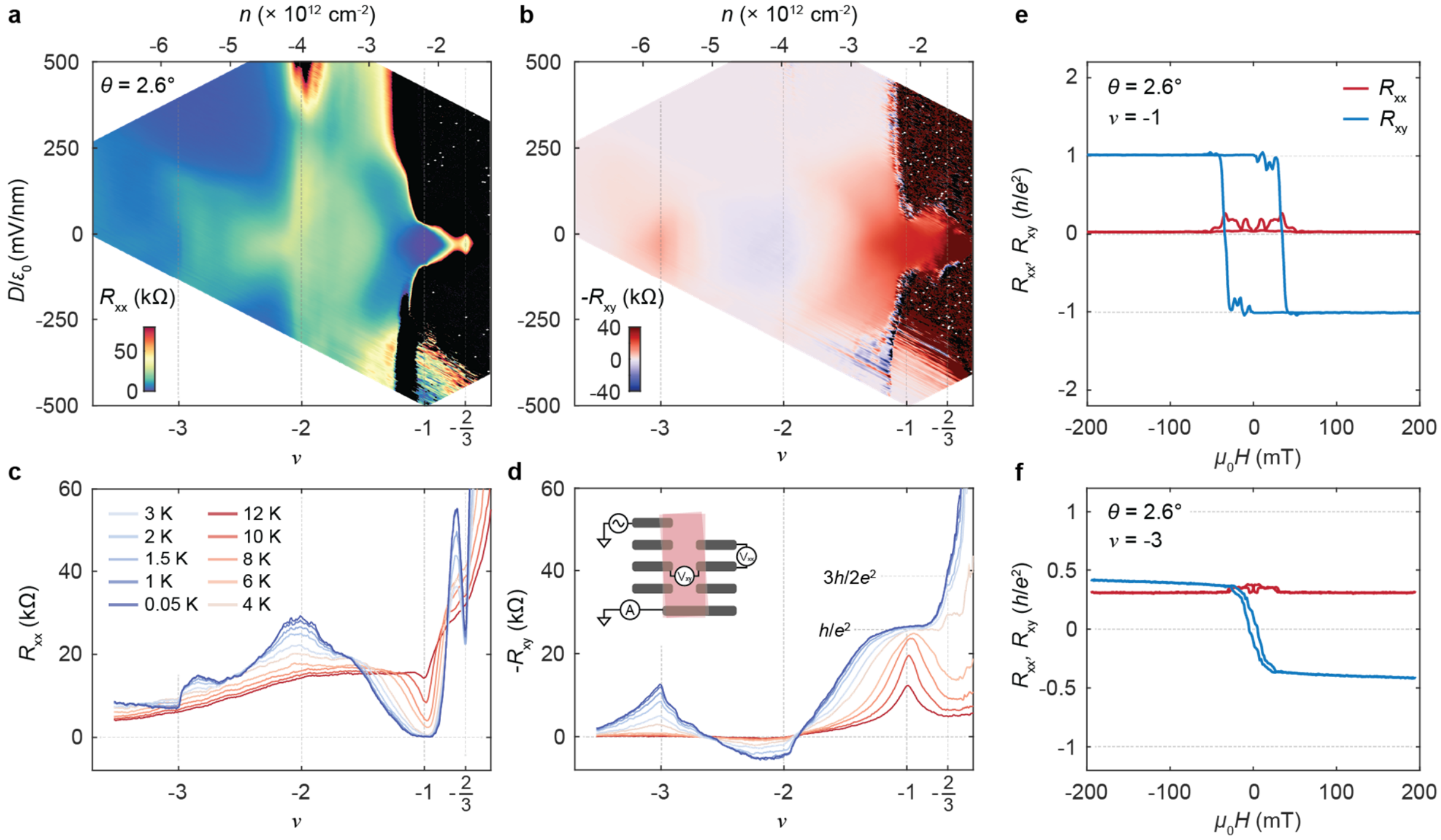

**Figure 2 | Transport measurements of anomalous Hall Effect near $\nu$ = -3 for 2.6° $tMoTe_2$.** **a**, **b**, Longitudinal ($R_{xx}$) resistance (**a**) and Hall ($-R_{xy}$) resistance (**b**) as a function of filling factor ($\nu$) and electric field ($D/\varepsilon_0$) at 15 mK. The carrier density ($n$) is shown on the top axis. $R_{xx}$ and $R_{xy}$ are symmetrized and anti-symmetrized at $|\mu_0 H| = 100$ mT. The resistance in the regions shaded in black cannot be reliably probed due to their highly insulating nature. **c**, **d**, Symmetrized $R_{xx}$ (**c**) and antisymmetrized $-R_{xy}$ (**d**) at $|\mu_0 H|$ = 500 mT as a function of $\nu$ at different temperatures. The electric field is kept at $D/\varepsilon_0 = 0$. $R_{xx}$ displays a local minimum at the integer and fractional Chern insulator states at $\nu$ = -1 and $\nu$ = -2/3, while $R_{xy}$ is quantized to $-h/e^2$ and $-3h/2e^2$, respectively. At $\nu$ = -3, $R_{xx}$ also dips while $R_{xy}$ peaks but not quantized, a signature of an incipient Chern insulator. Inset, contact scheme for measuring $R_{xx}$ and $R_{xy}$. **e**, **f**, Magnetic field dependence of $R_{xx}$ (red) and $R_{xy}$ (blue) at $\nu$ = -1 (**e**) and $\nu$ = -3 (**f**) at 15 mK.

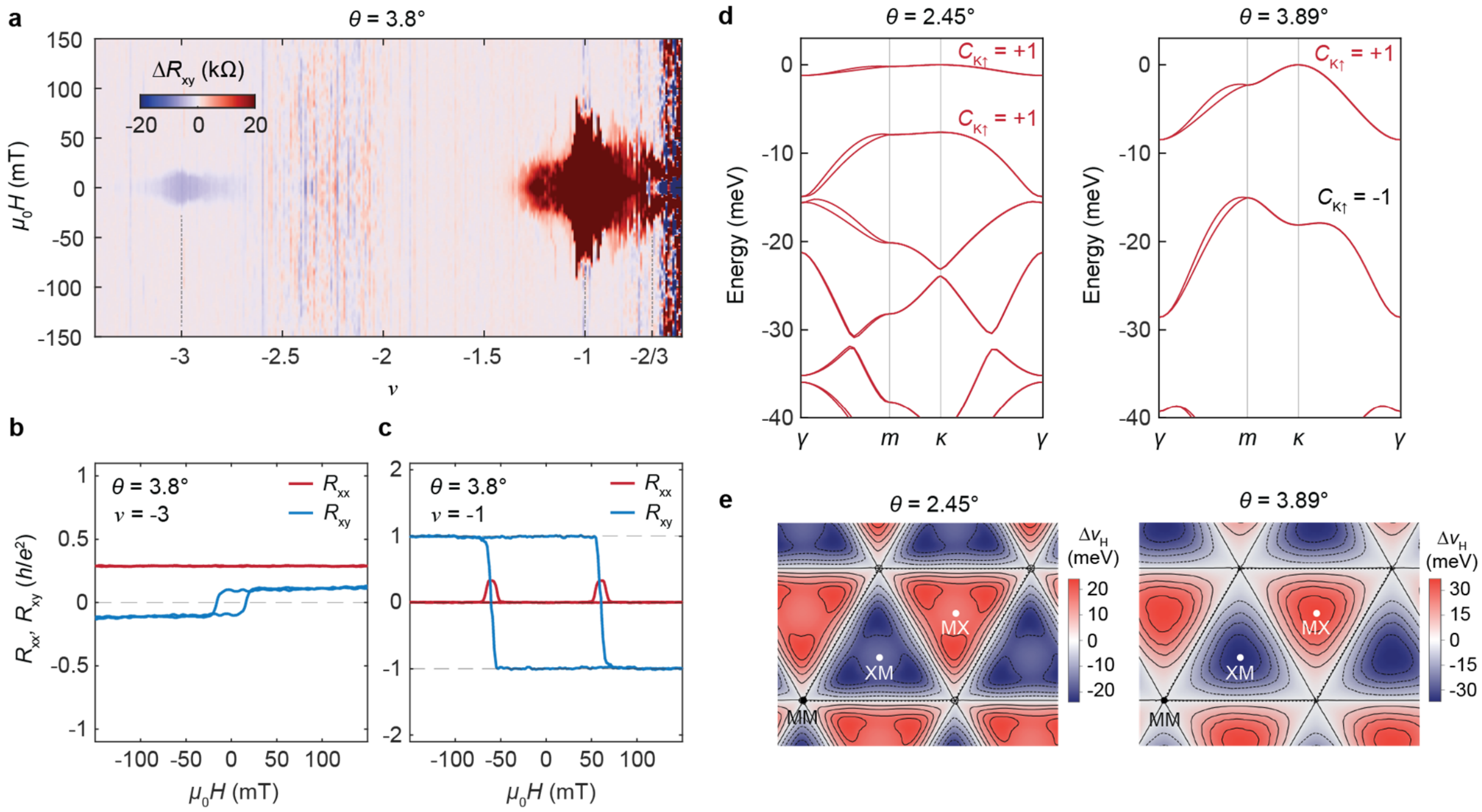

**Figure 3 | Twist angle dependent topology of the second Chern band. a**, The hysteretic component of $R_{xy}$ (difference between $R_{xy}$ for magnetic field swept up and down) versus $\nu$ for a 3.8° $tMoTe_2$ device at 15 mK. **b**, **c**, Magnetic field dependence of $R_{xx}$ (red) and $R_{xy}$ (blue) at 15 mK for the filling factor of $\nu$ = -3 (**b**) and $\nu$ = -1 (**c**) at $D/\varepsilon_0 = 0$ as the field is swept up and down. The sign of anomalous Hall effect signal at $\nu$ = -3 is opposite to that of the QAH state at $\nu$ = -1, as well as at $\nu$ = -3 in the 2.6° $tMoTe_2$ device, implying the opposite sign of Chern number. **d**, Large-scale density functional theory calculations of $tMoTe_2$ at twist angles of 2.45° (left) and 3.89° (right). While the Chern number of the first moiré miniband at the $K$ valley is both $C = +1$, the Chern number of the second miniband is given as $C = +1$ and $C = -1$ for 2.45° and 3.89°, respectively. **e**, Electrostatic moiré surface potential ($\Delta v_H$) for the two different twist angles. $\Delta v_H$ is obtained by calculating the difference in Hartree potential between the top layer and bottom layer in $tMoTe_2$. As the twist angle decreases the moiré orbitals are redistributed from the MX and XM sites towards the MM sites, causing a sign flip in the Chern number of the second moiré band.

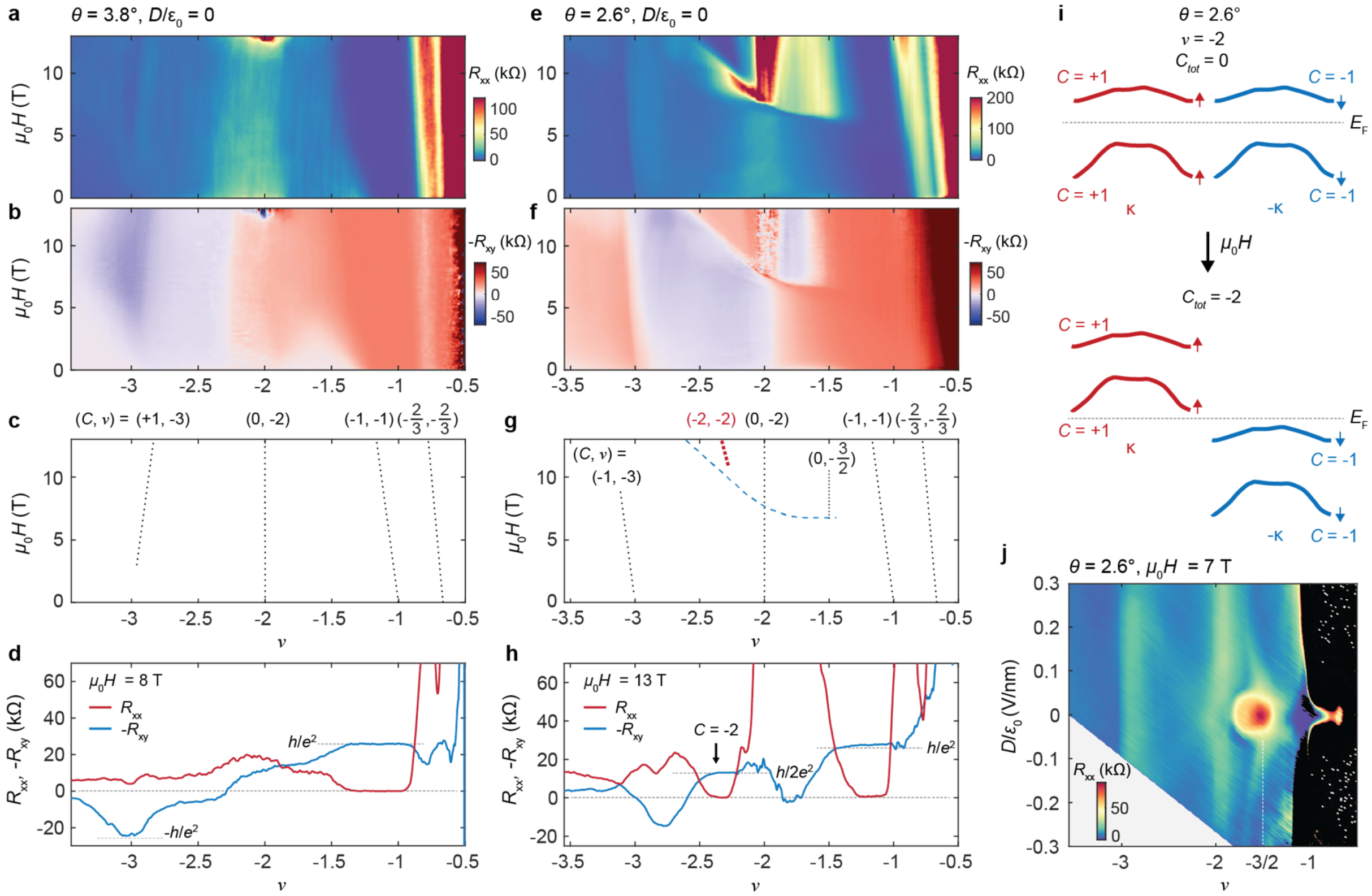

**Figure 4 | Topological band inversion at high magnetic field for $\nu$ = -2. a**, **b**, Landau fan of symmetrized/antisymmetrized $R_{xx}$ (**a**) and $R_{xy}$ (**b**) at $T$ < 30 mK for a 3.8° twisted $MoTe_2$ device at $D/\varepsilon_0 = 0$. **c**, Wannier diagram identifying the features present in the Landau fan. The Chern number ($C$) and intercept ($\nu$) are denoted. **d**, Linecuts of $R_{xx}$ and -$R_{xy}$ as a function of filling factor at a high field of 8 T, which shows $R_{xy}$ nearly quantized to $h/e^2$ at $\nu$ = -3, with an opposite sign to the $\nu$ = -1 QAH state. **e-g**, Similar to (a-c) but obtained from a 2.6° twisted $MoTe_2$ device. The blue dashed line in (**g**) identifies the phase transition at high field and the red dotted line corresponds to the $C$ = -2 state. **h,** Linecut of $R_{xx}$ and -$R_{xy}$ as a function of filling factor at a high field of 13 T. We identify a $C$ = -2 state near $\nu$ = -2, where $R_{xx}$ nearly vanishes and $R_{xy}$ is quantized to -$h/2e^2$. **i**, Schematic of the band inversion near $\nu$ = -2 for the 2.6° device. At zero field, the two valleys are equally populated resulting in a total Chern number of zero. At high fields, Zeeman shifts result in a band inversion between the first -$K$ band and second +$K$ band. Thus, the total Chern number becomes $C$ = -2 for $\nu$ = -2. For the 3.8° device, we also expect a similar band inversion, but the total Chern number remains $C$ = 0 for $\nu$ = -2. **j**, Symmetrized $R_{xx}$ as a function of $\nu$ and $D/\varepsilon_0$ at $\mu_0 H$ = 7 T for the 2.6° device. A signature of a resistive state at $\nu$ = -3/2 near $D/\varepsilon_0 = 0$ is visible, which does not disperse with magnetic field as seen in (g).

# Extended Data Figures for

**Ferromagnetism and Topology of the Higher Flat Band in a Fractional Chern Insulator**

Heonjoon Park[1†], Jiaqi Cai[1†], Eric Anderson[1†], Xiao-Wei Zhang[2†], Xiaoyu Liu[2], William Holtzmann[1], Weijie Li[1], Chong Wang[2], Chaowei Hu[1], Yuzhou Zhao[1,2], Takashi Taniguchi[3], Kenji Watanabe[4], Jihui Yang[2], David Cobden[1], Jiun-haw Chu[1], Nicolas Regnault[5,6], B. Andrei Bernevig[5,7,8], Liang Fu[9], Ting Cao[2], Di Xiao[1,2*], Xiaodong Xu[1,2*]

[1]Department of Physics, University of Washington, Seattle, WA, USA

[2]Department of Materials Science and Engineering, University of Washington, Seattle, Washington 98195, USA

[3]Research Center for Materials Nanoarchitectonics, National Institute for Materials Science, 1-1 Namiki, Tsukuba 305-0044, Japan

[4]Research Center for Electronic and Optical Materials, National Institute for Materials Science, 1-1 Namiki, Tsukuba 305-0044, Japan

[5]Department of Physics, Princeton University, Princeton, New Jersey 08544, USA

[6]Laboratoire de Physique de l'Ecole Normale Supérieure, ENS, Université PSL, CNRS, Sorbonne Université, Université Paris-Diderot, Sorbonne Paris Cité, 75005 Paris, France

[7]Donostia International Physics Center, P. Manuel de Lardizabal 4, 20018 Donostia-San Sebastian, Spain

[8]IKERBASQUE, Basque Foundation for Science, Bilbao, Spain

[9]Department of Physics, Massachusetts Institute of Technology, Cambridge, Massachusetts 02139, USA

† These authors contributed equally to the work.

Correspondence to: dixiao@uw.edu, xuxd@uw.edu

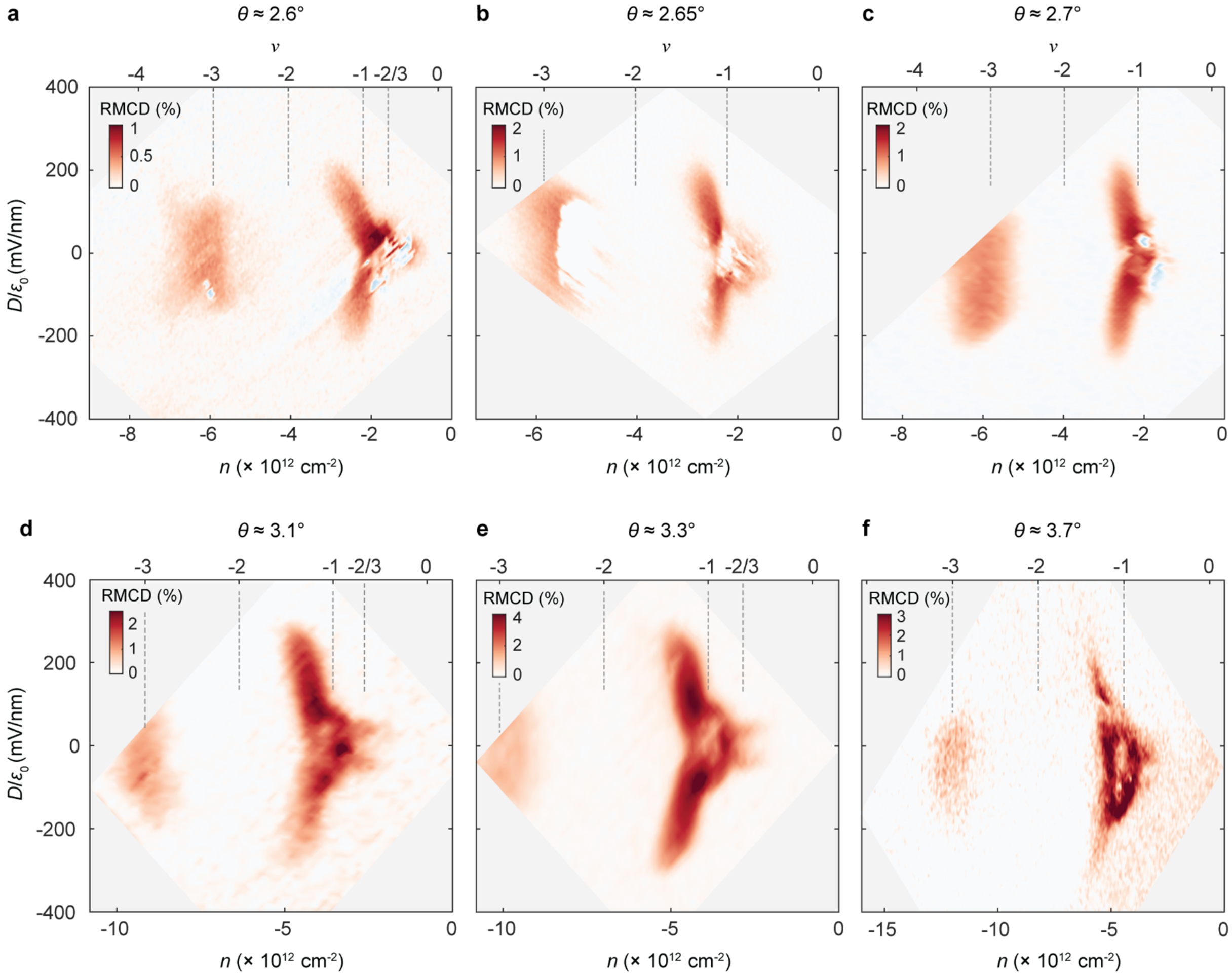

**Extended Data Figure 1 | Reproducibility of ferromagnetism at $\nu$ = -3 in devices with different twist angles. a-f**, Reflective magnetic circular dichroism (RMCD) signal as a function of carrier density ($n$, bottom axis) or filling factor ($\nu$, top axis) and displacement field ($D/\varepsilon_0$). Twist angles are indicated on top of each panel. All data were taken at zero magnetic field at a temperature of 1.6 K. Magnetic fluctuations are visible at $\nu$ = -1 and -3, which can be stabilized with a small out of plane field as seen in Fig. 1b in the main text.

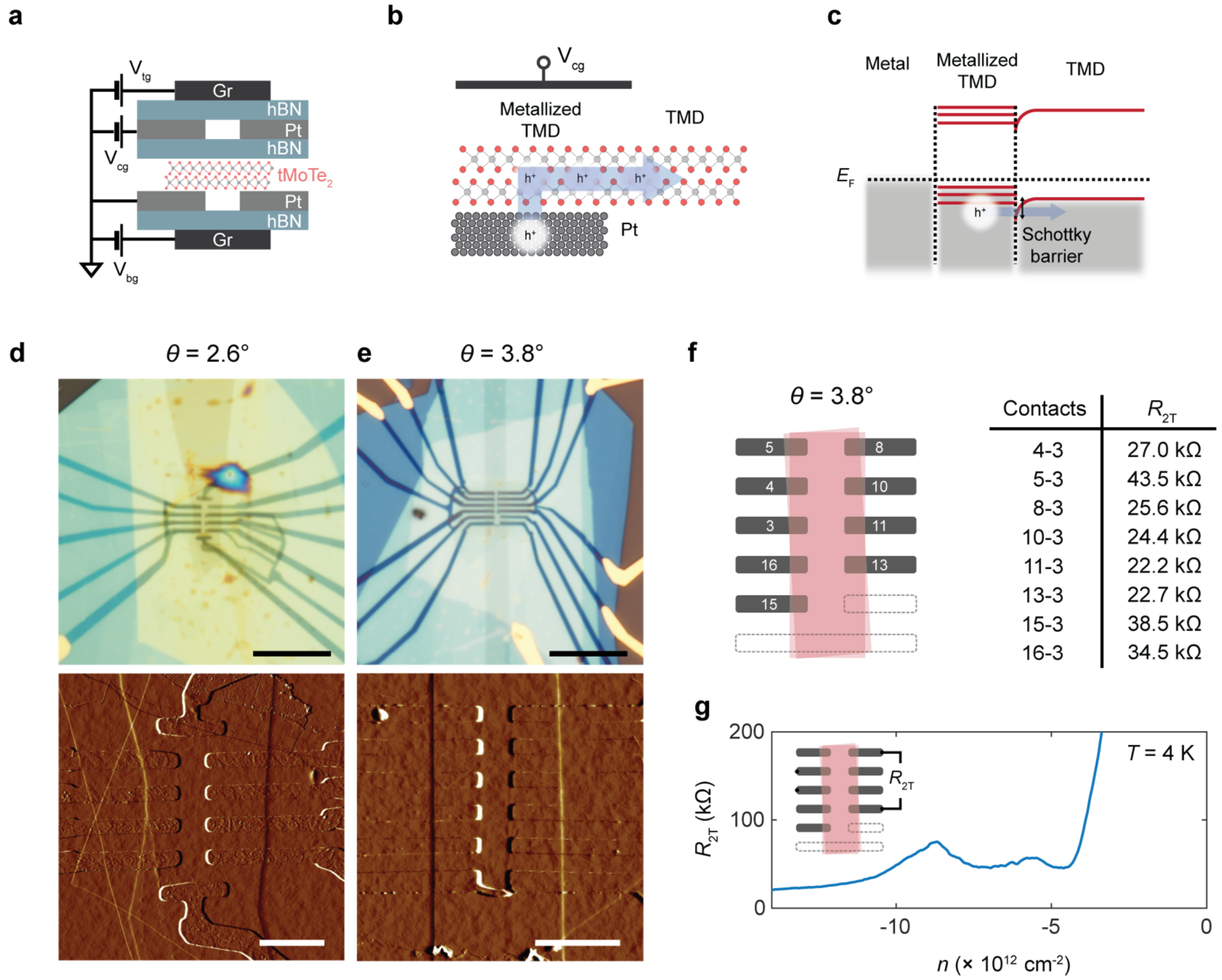

**Extended Data Figure 2 | Device images and contact resistance characterization. a,** Schematic of device structure. The top ($V_{tg}$) and bottom gates ($V_{bg}$) form a dual-gate geometry to independently control carrier density and electric field, while contact gates ($V_{cg}$) lower the contact resistance. **b,** Schematic of the contact region of Pt and transition metal dichalcogenide (TMD) interface. The hole travels sequentially from the Pt to the metallized TMD to the intrinsic TMD Hall bar region. **c**, Simplified energy level diagram of the three regions. The metallized TMD region forms defect-like localized bands as it is heavily strained by the contacts with a height of ~8 nm. A Schottky barrier forms between the metallized TMD and the intrinsic TMD, that can be overcome by heavily hole-doping the metallized TMD region with a large negative voltage on the contact gate. **d**, Optical image (top) of the 2.6° twisted $MoTe_2$ device and the AFM image (bottom) taken after transfer of the twisted $MoTe_2$ bilayer. The scale bars are 10 μm and 2 μm, respectively. **e**, Same for the 3.8° twisted $MoTe_2$ device. **f**, Characterization of $R_{2T}$ for different contacts within the 3.8° device, measured at the gate limit voltages of $V_{tg}$ = -8.4 V, $V_{bg}$ = -8 V, $V_{cg}$ = -3.1 V at $T$ = 4 K under an AC excitation of 0.1 mV. **g**, $R_{2T}$ as a function of carrier density ($n$) at $D/\varepsilon_0$ = 0 and $T$ = 4 K. The contact resistance starts to increase at very low doping levels. Inset, contact scheme for measuring $R_{2T}$.

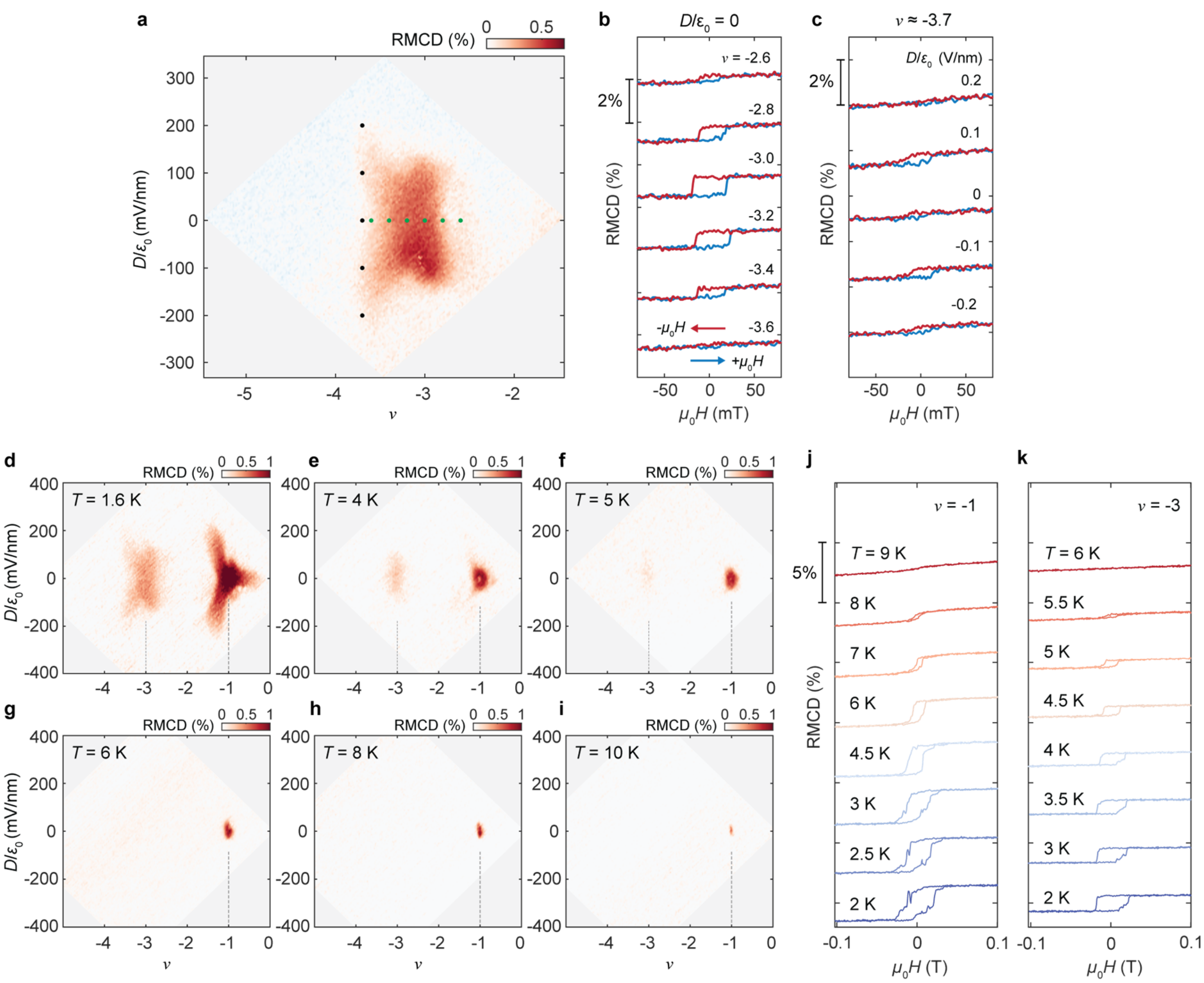

**Extended Data Figure 3 | Additional RMCD measurements of the 2.6° $tMoTe_2$. a,** RMCD signal as a function of filling factor ($\nu$) and electric field ($D/\varepsilon_0$) near $\nu = -3$, measured at zero magnetic field. A clear ferromagnetic 'wing-like' feature is visible. **b**, Magnetic field dependence of RMCD at selected filling factors, represented by the green dots in **a**. The electric field was fixed at zero ($D/\varepsilon_0 = 0$). **c**, Magnetic field dependence of RMCD at selected electric fields, marked by the black dots in **a**, at a fixed filling factor of $\nu \approx -3.7$. The data in **a-c** are taken at 1.6 K. **d-i**, RMCD maps as a function of $\nu$ and $D/\varepsilon_0$ at different temperatures ($T$), measured at a small field of 50 mT. The ferromagnetic signal at $\nu = -3$ disappears first around 6 K while the $\nu = -1$ signal persists up to approximately 9-10 K. **j**, **k**, Magnetic field dependence of RMCD at $D/\varepsilon_0 = 0$ for $\nu = -1$ (**j**) and $\nu = -3$ (**k**) at different temperatures from 2 K to 9 K (**j**) and 2 K to 6 K(**k**).

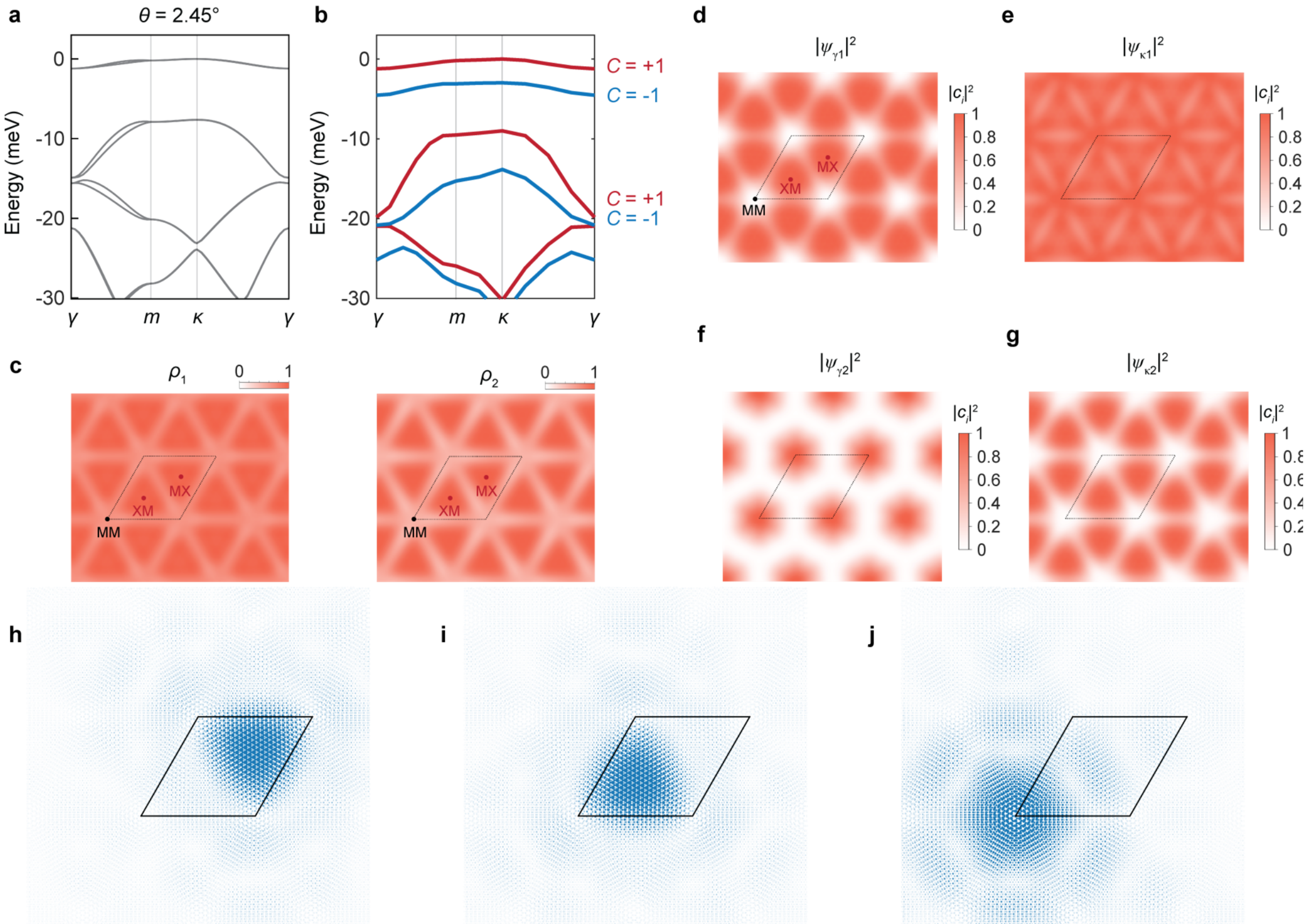

**Extended Data Figure 4 | Hartree-Fock bands and real-space wavefunction calculations.** **a,** Non-interacting single particle band structure for $tMoTe_2$ with a twist angle of 2.45°. **b**, Hartree-Fock band structure with spin-split bands for a small interaction strength of ε = 60. The Fock energy is inhomogeneous across momentum space, resulting in momentum dependent gaps across the Brillouin zone. It is noteworthy that the first moiré Chern band is fully gapped out, but the second Chern band remains gapless, consistent with transport measurements of a weak metallic state at $\nu = -3$. **c**, Real-space hole density of the first ($\rho_1$) and second ($\rho_2$) Chern bands calculated for 2.45° twisted $MoTe_2$. Both $\rho_1$ and $\rho_2$ form a honeycomb lattice at the MX and XM sites, leading to ferromagnetism via direct exchange. **d**, **e**, Wavefunction distribution across the moiré unit cell for the first Chern band at the γ point (**d**) and κ/κ′ point (**e**). Both form a honeycomb lattice, consistent with the hole density distribution in **c**. **f**, **g**, Wavefunction distribution for the second Chern band at the γ point (**f**) and κ/κ′ point (**g**). The wavefunctions are normalized with respect to their maximum value within the moiré unit cell. In contrast to the κ/κ′ point, which remains a honeycomb lattice, the γ point forms a triangular lattice which has decreased mean-field Fock energy. As a result, the gap between the two spin-split Chern bands in (**b)** is less for the γ point compared to the κ/κ′ point. **h**, **i**, **j**, Wannier orbitals for each valley (spin) localized at the MX, XM and MM site, respectively.

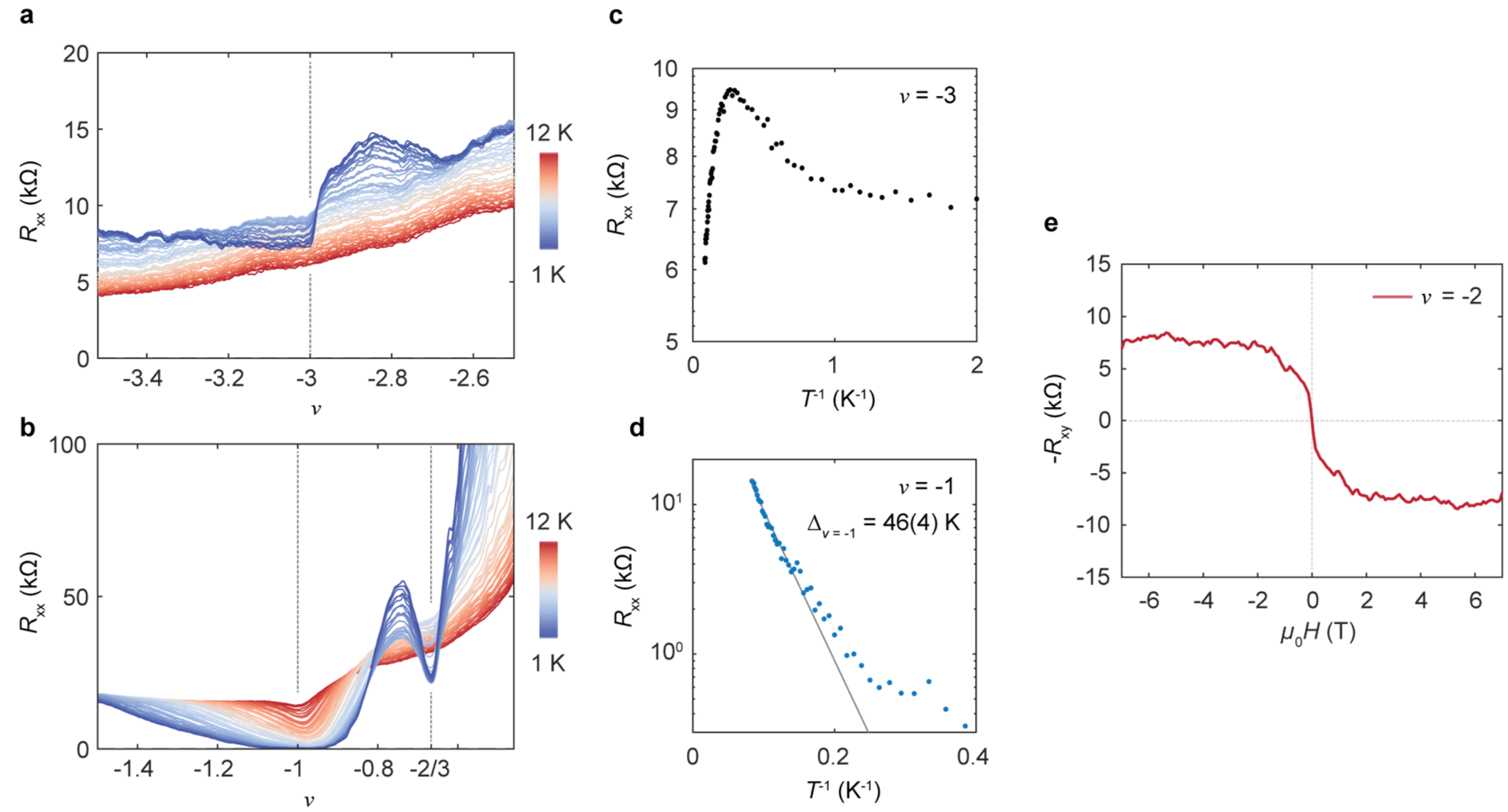

**Extended Data Figure 5 | Temperature dependence of $R_{xx}$ and gap extraction. a-b,** $R_{xx}$ versus $\nu$ near $\nu$ = -3 (**a**) and $\nu$ = -1 (**b**), respectively, at selected temperatures from 1K to 12K. **c-d**, Extracted $R_{xx}$ as a function of inverse temperature ($T^{-1}$) at $\nu$ = -3 (**c**) and $\nu$ = -1 (**d**). The thermal activation gap is found to be 46(4) K for $\nu$ = -1, while it appears that $\nu$ = -3 is not fully gapped. **e**, $R_{xy}$ as a function of magnetic field near $\nu$ = -2 and $D/\varepsilon_0 = 0$, measured at $T$ <100 mK.

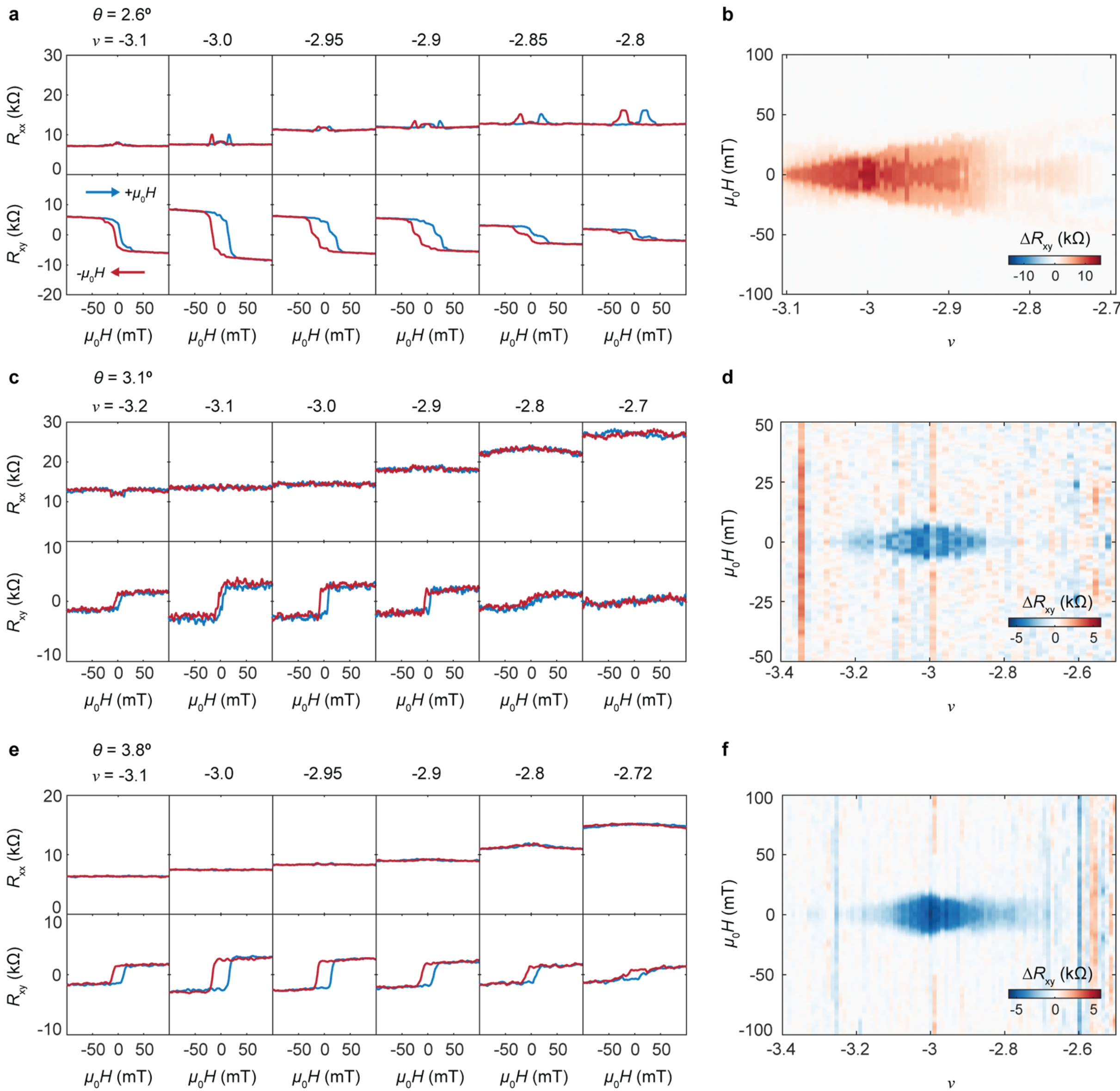

**Extended Data Figure 6 | Filling factor dependent anomalous Hall effect near $\nu$ = -3 for different twist angles. a,** Magnetic field dependence of $R_{xx}$ (top) and $R_{xy}$ (bottom) at selected filling factors near $\nu$ = -3 for 2.6° twisted $MoTe_2$ device. **b**, $\Delta R_{xy}$ versus magnetic field and filling factor. Here, $\Delta R_{xy}$ is obtained by taking the difference of $R_{xy}$ between sweeping magnetic field up and down. The data in **a** and **b** are taken at $T$ = 15 mK and $D/\varepsilon_0$ = 0 from a different cool down of the same sample in the main Figure 2. **c**, **d**, Similar data for a 3.1° twisted $MoTe_2$ device, and **e**, **f**, similar data for a 3.8° twisted $MoTe_2$ device. The $\Delta R_{xy}$ is positive (red) at the $\nu$ = -1 Chern insulator state ($C$ = -1) for all devices. A sign reversal of the AHE at $\nu$ = -3 is evident between the 2.6° device and the devices with twist angles of 3.1° and 3.8°.

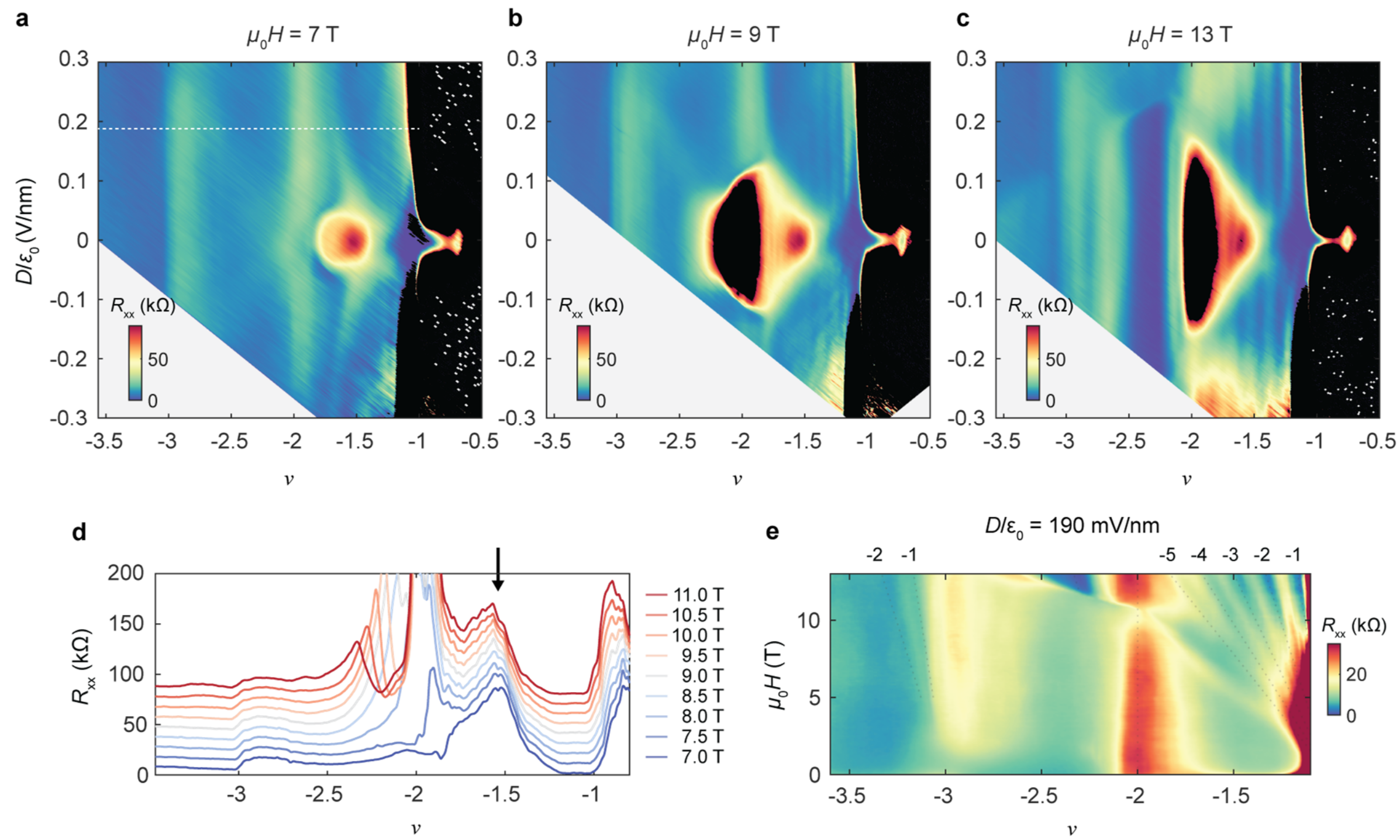

**Extended Data Figure 7 | Dual gate maps at high magnetic fields and Landau fan at high electric field for filling factor assignment of the 2.6° device. a, b, c**, Longitudinal resistance $R_{xx}$ as a function of filling factor $\nu$ and electric field $D/\varepsilon_0$ at out-of-plane magnetic fields ($\mu_0H$) of 7 T (**a**), 9 T (**b**), and 13T (**c**). The regions in black are inaccessible due to their high resistance. A resistive state at $\nu$ = -3/2 is visible near $D/\varepsilon_0$ = 0. **d**, Linecuts of $R_{xx}$ at selected magnetic fields. Each curve is displaced by 10 kΩ for clarity and the arrow highlights the non-dispersing resistive maximum at the $\nu$ = -3/2 state. **e**, The Landau fan of $R_{xx}$ measured as a function of $\nu$ and $\mu_0H$ at a finite electric field of $D/\varepsilon_0$ = 190 mV/nm, denoted as the white line in **a**. Clear quantum oscillations are visible stemming from the filling factor of $\nu$ = -3 and $\nu$ = -1, which allows accurate filling factor assignment for the main figures. The degeneracy of the Landau levels is denoted on the top axis. Data are taken at a temperature of 15 mK.

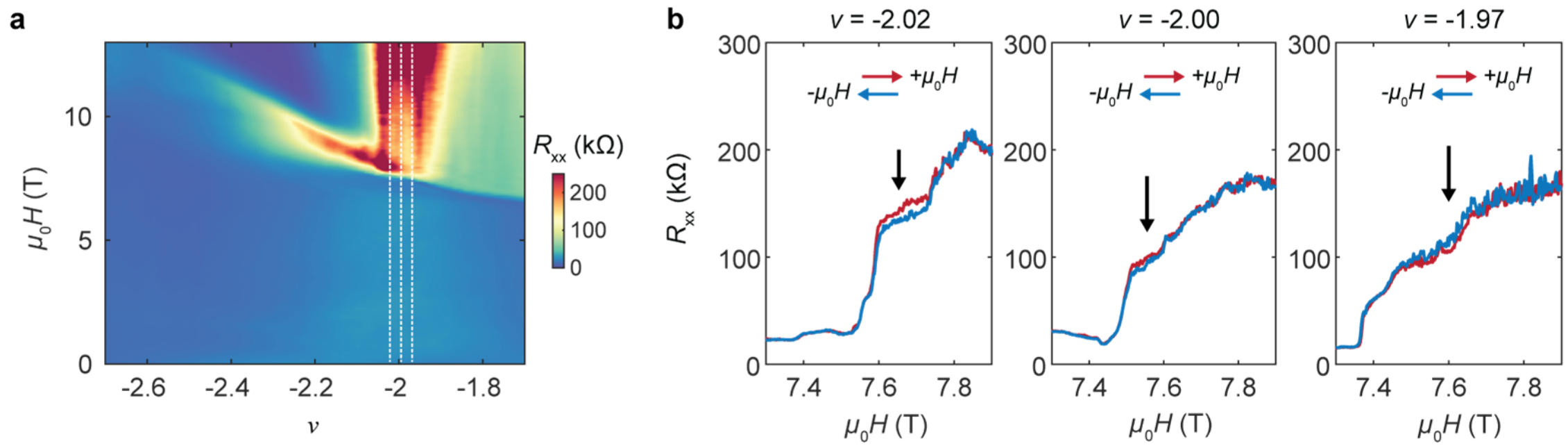

**Extended Data Figure 8 | Magnetoresistance and first order phase transition near $\nu$ = -2.** **a**, Zoom in of the Landau fan diagram in the main Figure 4e near $\nu$ = -2. **b**, Magnetoresistance as a function of sweeping the field up and down near the filling factor of near $\nu$ = -2. A clear hysteresis can be seen, which is a signature of a first order spin-flip transition due to Zeeman splitting.

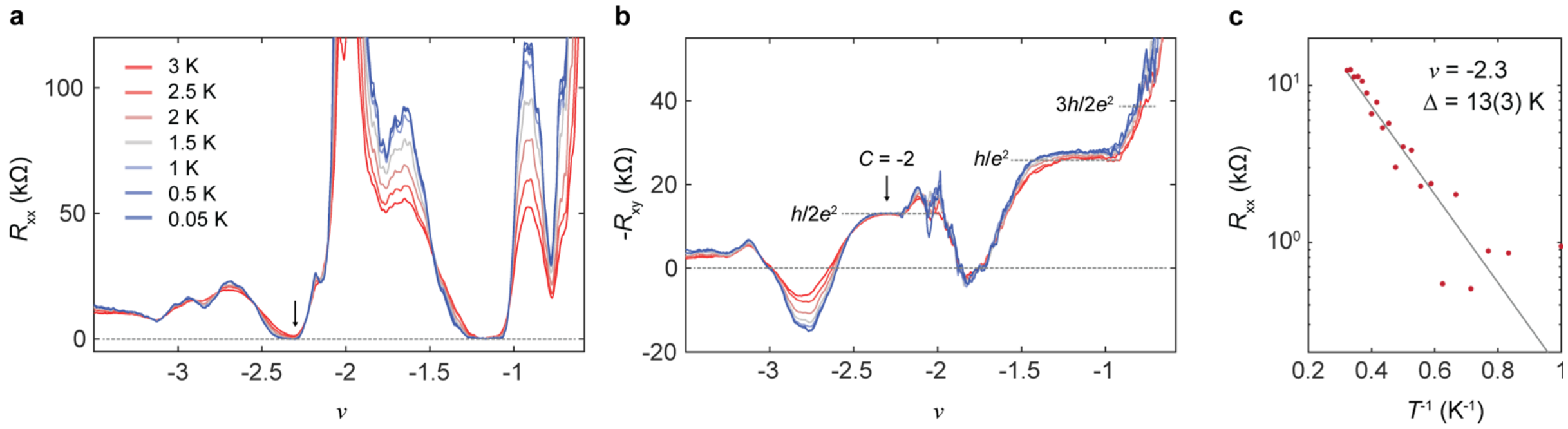

**Extended Data Figure 9 | Temperature dependence of $C$ = -2 Chern insulator state at high magnetic fields. a, b**, Filling factor dependent $R_{xx}$ (**a**) and $R_{xy}$ (**b**) at selected temperatures at a fixed field of 13 T. Data are symmetrized and anti-symmetrized at ±13 T. Note that at low doping levels the contact resistance does not become negligible, hence the deviation from FCI behavior. **c**, Thermal activation behavior of $R_{xx}$ near the filling factor of $\nu$ = -2.3, which indicates an energy gap of 13(3) K.